\documentstyle[proceedings,numreferences,epsfig,citesort]{crckapb}

\begin{opening}
\title{
A SIMPLE METHOD TO MIX GRANULAR MATERIALS}

\author{SEAN McNAMARA}
\institute{Levich Institute, Steinman Hall T-1M, \\
    140th St and Convent Ave, \\
    New York, NY 10031, USA \\
    Correspondence to: \verb|mcnamara@levdec.engr.ccny.cuny.edu| }
\author{STEFAN LUDING}
\institute{Institute for Computer Applications 1\\
           Pfaffenwaldring 27, 70569 Stuttgart, GERMANY}

\end{opening}

\runningtitle{A SIMPLE METHOD TO MIX GRANULAR MATERIALS}

\begin{document}

\begin{abstract}
We show that a mixture of two species of granular particles with equal
sizes but differing densities can be either segregated or mixed by
adjusting the granular temperature gradient and the magnitude of the
gravitational force.  In the absence of gravity, the dense, heavy particles
move to the colder regions.  If the temperature gradient is put into a
gravitational field with the colder regions above the hotter, a uniform
mixture of light and heavy particles can be attained.  This situation can
be realized in a container of finite height with a vibrating bottom, placed
in a gravitational field.  We present a relation between the height of the
container, the particle properties, and the strength of gravity required to 
minimize segregation.
\end{abstract}
~\vspace{-1.5cm}~\\

\section{Introduction}

Segregation and mixing of granular material is of eminent importance 
for industrial operations and it has been subject to research since decades.
However, both effects are not yet completely understood 
and thus cannot be controlled under all circumstances. 
Traditional experimental methods and theoretical
approaches are nicely complemented by numerical simulations which 
in the last few years have developed tremendously \cite{herrmann98}. 
For a review which covers a broad practical experience of
segregation see Ref. \cite{chowhan95} and references therein. 

Segregation can be driven by geometric effects, shear, percolation 
and also by a convective motion of the small particles in the
system \cite{knight93}. In vibrated systems, the segregation due to 
convection appears to be orders of magnitude faster than segregation due 
to purely geometrical effects \cite{duran93,duran94}. 
In rotating drums, another archetype of many industrial
devices, several segregation processes acting in parallel are reported
\cite{cooke76}; in three-dimensional devices, axial and
longitudinal segregation are observed \cite{oyama39,gupta91,nakagawa94}
simultaneously.  For axial segregation, particle percolation is
reported to be responsible \cite{gupta91}, while longitudinal
segregation is related to different surface flow properties in the cylinder
\cite{zik94,hill94}. 

\def\ave#1{\langle #1 \rangle}

\begin{figure}[hbt]
 \begin{center}
 \epsfig{file=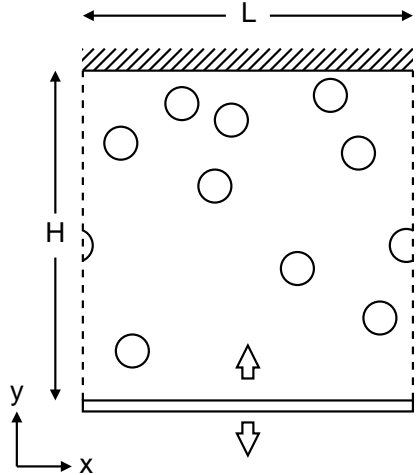,width=6.0cm,clip=,angle=0}
\end{center}
\caption{A sketch of the studied system}
\label{fig:system}
\end{figure}

In this paper, we investigate a model segregation problem which suggests
a simple way to obtain uniform mixtures of two species.  We show a
sketch of the system in Fig.~\ref{fig:system}.

$N$ particles are placed in a
container of height $H$ in the presence of gravity. 
The side walls have been replaced with periodic boundary conditions. 
Energy is supplied
to the container by vibrating the bottom using a symmetric sawtooth wave
with velocity $V$.  The top wall is stationary.  $N_A$ of
the particles have mass $m_A$, and the rest have mass $m_B$.  We will
take $m_A>m_B$.  Though the particles have different mass, they all have
the same radius $a$.  We model the loss of energy during collisions with a
restitution coefficient $r<1$.  Our detailed study of a similar system
with identical particles is Ref.~\cite{mcnamara98b}.

\section{The Mixing Mechanism}

\begin{figure}[ht]
\begin{center}
{\large (a) \hfill (b)} \vspace{-.45cm}\\
 \epsfig{file=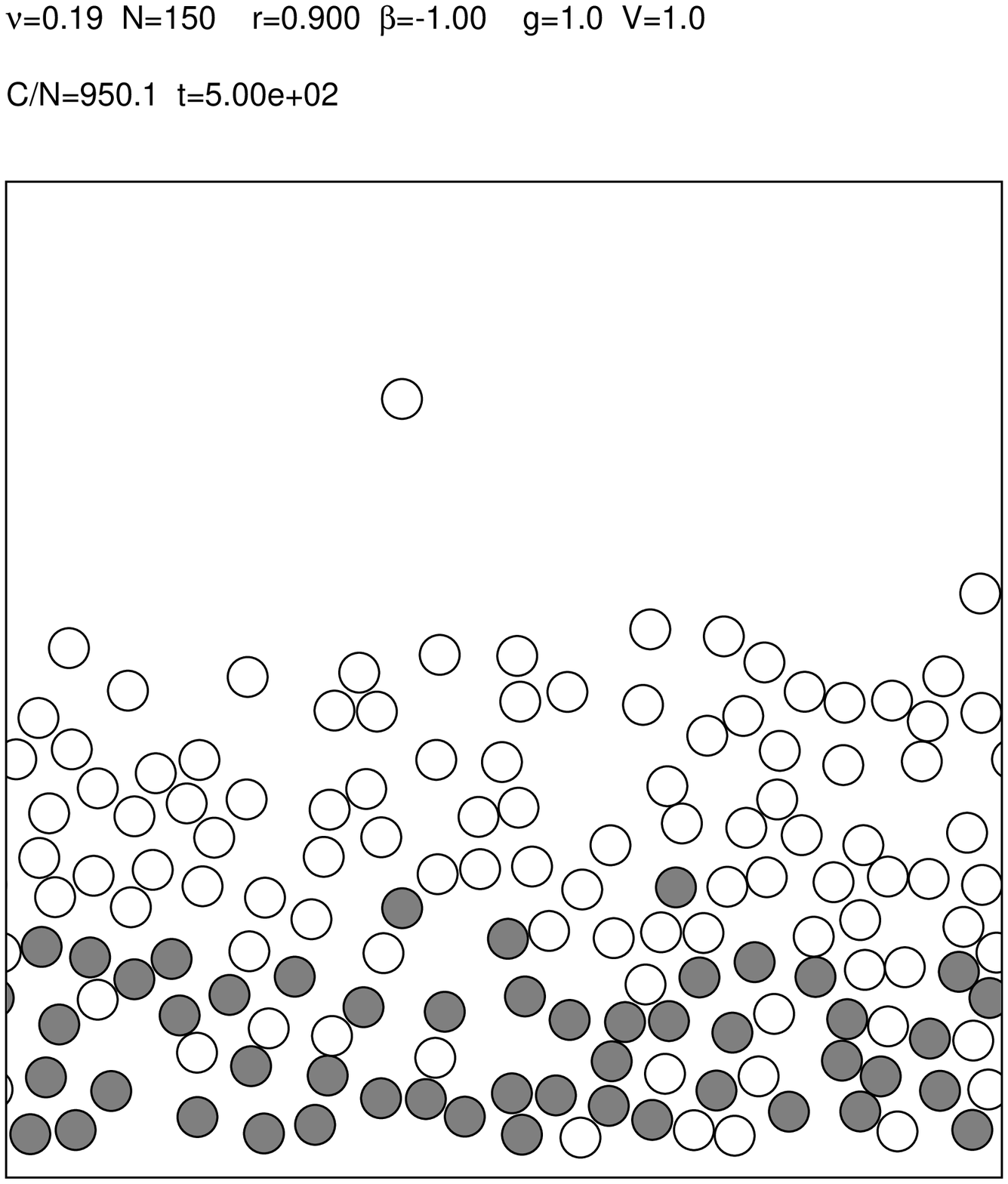,width=5.6cm,clip=,angle=0}
 \epsfig{file=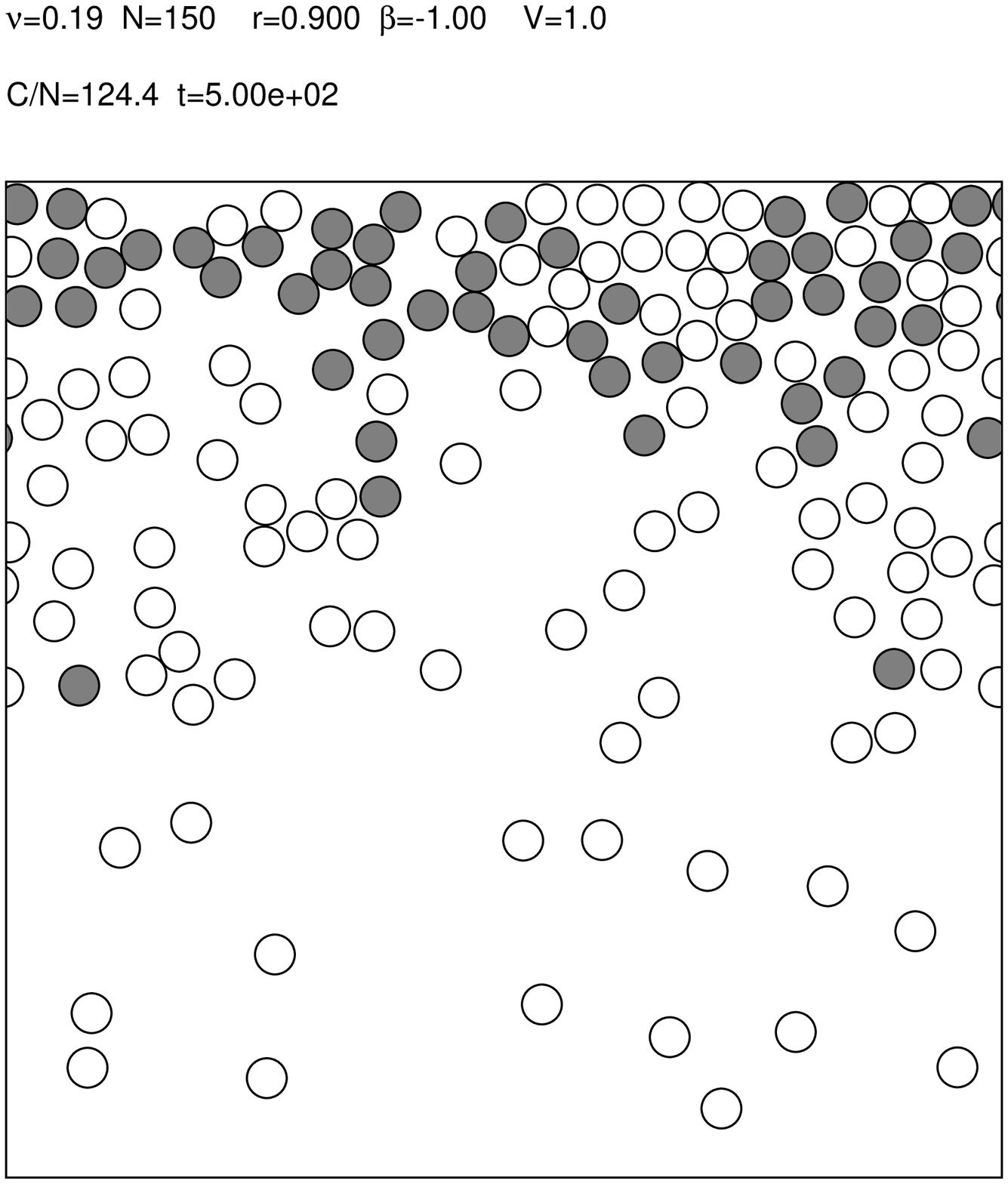,width=5.6cm,clip=,angle=0}~~~
\end{center}
\caption{(a) A simulation with $N=150$, $N_A=50$, $L=H=50$, $m_A/m_B=10$, 
$g=1$, $V=1$, $r=0.9$.  The dense particles (mass $m_A$) are shaded.
(b) same as (a), except $g=0$.}
\label{fig:seg1}
\end{figure}

We find that it is possible to obtain uniform mixtures of the two species
by pitting two segregation mechanisms against each other.  When the particles 
rarely touch the top of the container, all
the dense particles are found near the bottom of the plate
(see Fig.~\ref{fig:seg1}a).   A similar effect occurs in the upper atmosphere,
where different molecular species are sorted by weight \cite{jeans}.

\begin{figure}[ht]
\begin{center}
  \epsfig{file=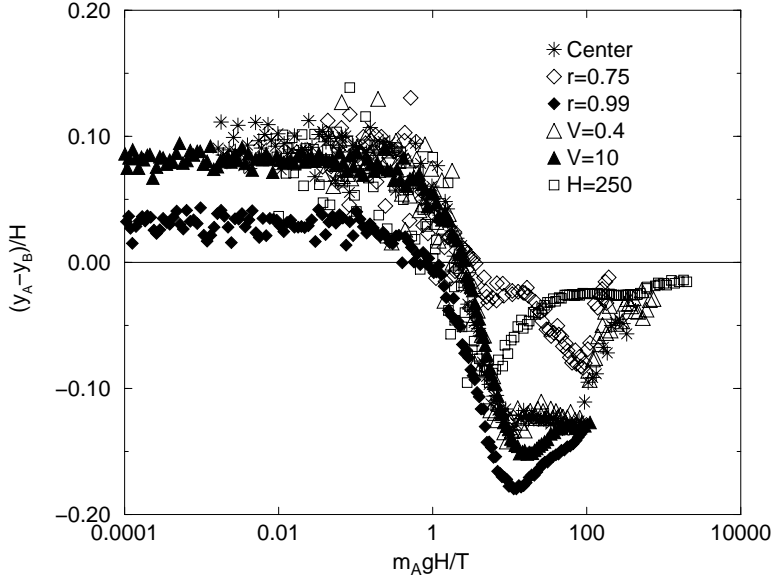,width=11cm,clip=,angle=0}
\vspace{-1cm}~\\
\end{center}
\caption{Segregation of the two species measured by the difference between
$y_A$, the center of mass of species $A$ and $y_B$, the
center of mass of species $B$.  This difference is plotted as a function
of $m_AgH/T$, where $T$ is the granular temperature, here defined as the
average kinetic energy per particle.
The points marked ``Center'' have $V=2$, $m_A/m_B=2$, $N_A=50$, $N_B=100$,
$H=L=50a$, and $r=0.95$, with $g$ swept over six orders of magnitude.  The
other points are the same, except for the parameter values marked on the
graph.  In all cases, the time unit is fixed by the wall vibration period,
and the mixture is equilibrated for $300$ time units, and then data
is averaged for $1500$ time units.}
\label{fig:yAyB}
\end{figure}

On the other hand, when gravity is turned off, the particles are pushed against
the top plate and the dense particles are found close to the upper plate
(Fig.~\ref{fig:seg1}b).
By smoothly varying between these two situations, it is possible
to obtain a situation where the two species are uniformly mixed.  In
Fig.~\ref{fig:yAyB}, we plot the difference between $y_A$, the center of mass of
the heavy particles, and $y_B$, the center of mass of the light
particles, normalized by the height of the container.
Mixing is optimal when the difference 
between the species' centers of mass vanishes.

We see that the state of maximum mixing is obtained near $m_AgH/T \sim 2$
for all values of the parameters, except for almost elastic particles 
($r=0.99$).  Here, $T$ is the granular temperature, defined as the
average kinetic energy per particle: $T\equiv (1/2N) \sum m_i v_i^2$.
When the ratio of both energies is near
unity, it means that the kinetic and potential energies of the particles
are comparable.  
The tendency of $(y_A-y_B)/H$ to approach $0$ for large gravities $m_AgH/T>10$
is due to the initial conditions.  Initially, all particles are arranged
in a lattice just above the vibrating floor.  Due to the
large gravity, it is very difficult for particles to change places,
and the mixture keeps its original configuration for a very long
time.  For $m_AgH/T<10$, the particles change places often, and
$(y_A-y_B)/H$ is independent of initial conditions.

\begin{figure}[ht]
\begin{center}
 \epsfig{file=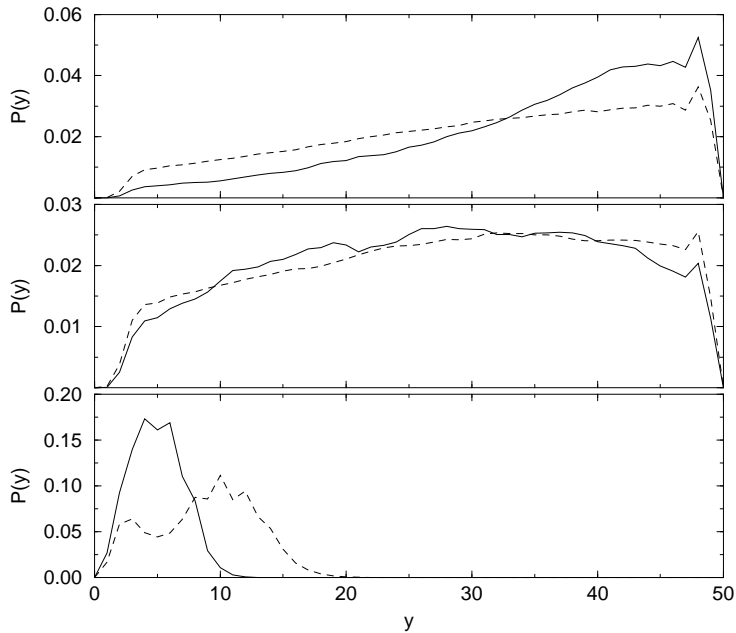,width=10cm,clip=,angle=0}
\end{center}
\caption{Profiles of the density of each species as a function of $y$.
The solid curve represents the heavy species with mass $m_A$, and
the dashed curve is the light species with $m_B$.
All simulations have $m_A/m_B=2$, $V=2$, $H=50$, $N_A=50$, $N_B=100$.
The top figure has $g=10^{-4}$, the middle has $g=0.3169$, and the
bottom has $g=10^2$.  The densities are given in terms of the probability
of a single particle of each species to have a given height; the area
under all curves integrates to $1$.}
\label{fig:gprofiles}
\end{figure}

To show more closely what happens with the densities of the different
species, we show in Fig.~\ref{fig:gprofiles} the concentrations of each 
as a function of height for three different simulations; 
one at small $g$, one at large $g$, and one where the particles are 
nicely mixed. In the situations with extremal $g$ values, we obtain
rather strong density gradients, while in the case of optimal mixing
the density gradients are small, i.e.~the density is almost constant
throughout the system.

\section{Discussion and Conclusion}

\begin{figure}[ht]
\begin{center}
{\large (a) \hfill (b)} \vspace{-.45cm}\\
 \epsfig{file=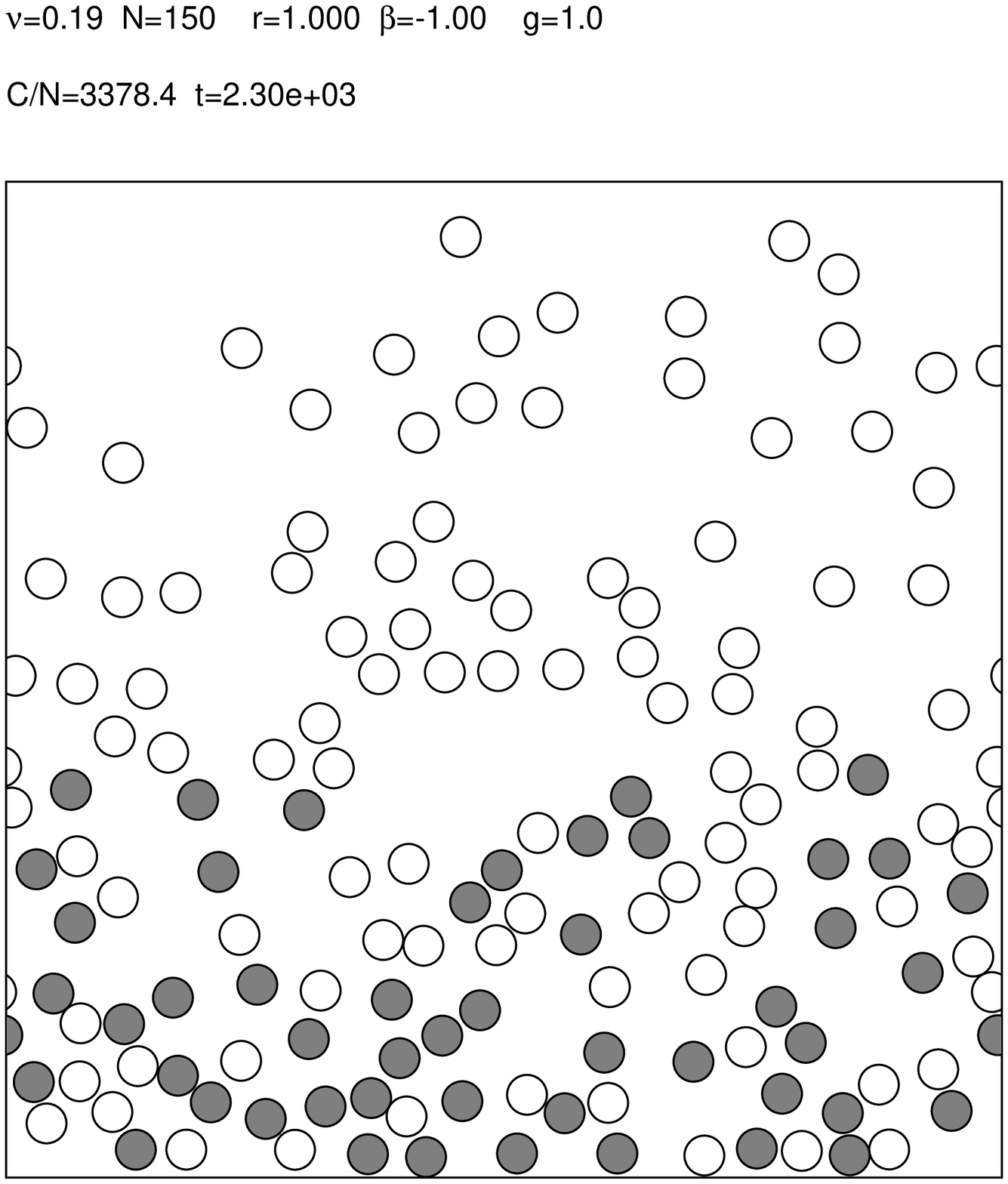,width=5.6cm,clip=,angle=0}
 \epsfig{file=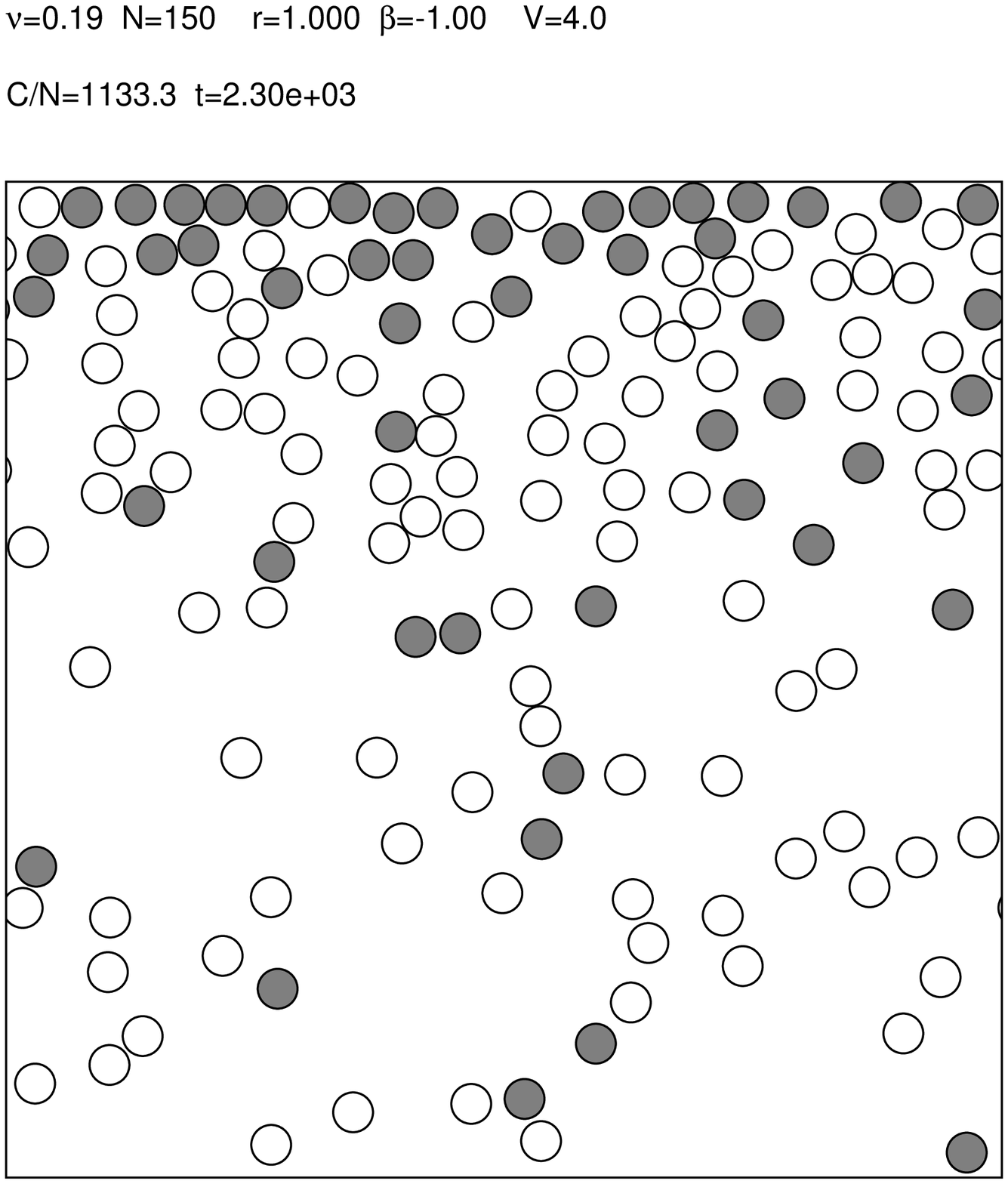,width=5.6cm,clip=,angle=0}~~~
\end{center}
\caption{(a) A simulation with $N=150$, $N_A=50$, $L=H=50a$, $m_A/m_B=2$,
$g=1$, $r=1.0$.  All the walls are stationary, so no energy is added
or subtracted.  Note that the dense particles sink to the bottom.
(b) same as except that $g=0$ and when a particle touches the upper or
lower wall, a new velocity is selected from a Maxwellian velocity
distribution with a certain temperature.  The temperature of the lower
wall is $160$ times as large as the temperature of the upper wall.  Smaller
temperature gradients also sort the particles by density, but it is much
less visually striking.}
\label{fig:r=1segs}
\end{figure}

Each of the two segregation mechanisms can be observed also with perfectly 
elastic (dissipationless) particles.  In Fig.~\ref{fig:r=1segs}(a), we show a 
binary gas of elastic particles under
gravity in the absence of forcing.  The heavy particles accumulate at
the bottom.  In Fig.~\ref{fig:r=1segs}(b), we show a binary gas in the
absence of gravity, subjected to a thermal gradient.  Now the particles
particles accumulate against the upper, cold wall.  Therefore, neither
segregation mechanism relies on the dissipation of energy during collisions.
This dissipation
serves only to set up the necessary gradients which drive the segregation
of the particles (see also the paper by Luding, Strau\ss{}, and
McNamara in this proceedings). 

To use this method to mix granular materials, the particles could be put
into a chamber like the one shown in Fig.~\ref{fig:system}.  To obtain
the proper value of $m_AgH/T$, it is perhaps most convenient to
adjust the height of the container $H$.  It is also possible to control
$T$ by changing the vibration velocity $V$ \cite{mcnamara98b}.
One possible disadvantage is that only a small amount of material can be
mixed at one time.  It also may be difficult in practice to adjust
$H$ or $T$ correctly.  Replacing the periodic boundaries with side
walls may also introduce new effects.

\section*{Acknowledgements}

Inspiring discussions with H.~J.~Herrmann are appreciated, and
we gratefully acknowledge the support of IUTAM, the National Science
Foundation and the Department of Energy. S.L. also thanks the Deutsche
Forschungsgemeinschaft, and S.M. the Alexander-von-Humboldt foundation,
and the Geosciences Research Program, Office of Basic Energy Sciences,
US Department of Energy.

\bibliographystyle{prsty}
\bibliography{/home/lui/LIT/granulates}

\end{document}